\documentclass[%
 reprint,
 amsmath,amssymb,
 prb,
]{revtex4-2}

\usepackage{graphicx}
\usepackage{dcolumn}
\usepackage{bm}
\usepackage{scrextend}
\usepackage{hyperref}
\usepackage{dcolumn}

\usepackage{xcolor}
\hypersetup{
    colorlinks,
    linkcolor={red!40!black!100!},
    citecolor={blue!30!black!100},
    urlcolor={blue!30!black!100}
}
\newcolumntype{S}{D{/}{/}{1}}
\begin{document}

\preprint{APS/123-QED}

\title{Electronic and optical properties of native point defects in CuInS$_2$ and CuGaS$_2$\\}

\author{Henry Phillip Fried}\email{henry.p.fried@gmail.com}
\author{Daniel Barragan-Yani}
\author{Ludger Wirtz}%
\affiliation{%
 Department of Physics and Material Science, University of Luxembourg \\
}%
\date{\today}

\begin{abstract}
We present a detailed study of common intrinsic defects in CuInS$_2$ and CuGaS$_2$ using the Heyd, Scuseria and Ernzerhof (HSE) hybrid functional scheme. The impact of the two HSE parameters, $\alpha$ and $\omega$  on the band gap and compliance with the generalized Koopmans' theorem is investigated. Using the formation energy formalism and calculated thermodynamic charge-transition levels, we assess the electronic properties of the defects and explore the connection of charge-transition levels with optical-transition levels. Calculated Franck-Condon shifts for emission highlight the importance of lattice relaxation for the attribution of defects to luminescence peaks. Our results show that once these effects are included, predictions become closer to photoluminescence measurements available in literature.

\end{abstract}

\maketitle


\section{\label{sec:Introduction} Introduction}
Defects are essential for many devices, such as transistors, solid-state quantum sensing, or solar cells~\cite{SemiconductorBook1, SemiconductorBook2, Queisser1998,Atature2018May,Zhang2020Sep, Rovny2024Dec}. However, impurities can also drastically reduce the performance of the device. For example, in solar cells they can cause unwanted doping compensation or create deep defects that act as recombination centers, thus reducing the efficiency of the device~\cite{SolarBook2016}. To successfully design the next generation of devices, we need a clear understanding of the physics of promising new materials and their defects.

In the context of new materials for photovoltaic applications, the family of chalcopyrites has been growing in recent years as promising low- and wide-band gap absorber materials within solar cells~\cite{CIGStandemsolarcell1, Spindler2019, Shukla2021a}. Most recently, CuInS$_2$ and CuGaS$_2$ have received attention because their quaternary alloy Cu(In,Ga)S$_2$ allows the adjustment of the band gap depending on the [Ga]/[Ga+In] content, making them suitable for tandem solar cell applications. To date, an efficiency of over 16.1\% has been demonstrated for Cu(In,Ga)S$_2$ (CIGS)-based solar cells~\cite{Adeleye2025Feb}. A fairly complete assignment of defect levels has been achieved for quaternary  Cu(In,Ga)Se$_2$ compounds~\cite{Spindler2019}; however, similar efforts for Cu(In,Ga)S$_2$ are still ongoing.

Photoluminescence (PL) spectroscopy is often used to investigate defects~\cite{CharacterizationBook1, CharacterizationBook2}. In the quaternary compound Cu(In,Ga)S$_2$, these measurements reveal two optical transitions associated with deep defects around 1.35 eV (D1) and 1.1 eV (D2), respectively~\cite{Shukla2021a}. Photoluminescence measurements on CuGaS$_2$ also show two deep defect transitions at approximately 2.15 eV (DD1) and 1.85 eV (DD2) ~\cite{Eberhardt2003Sep,Botha2007May,Adeleye2024Aug}. Moreover, the behavior of the DD1 peak is dependent on the [Cu]/[Ga] ratio. So far, only Cu$_\text{III}$ has been proposed as a candidate defect for the D1 peak~\cite{Shukla2021a}, while it has  been hypothesized that more than one defect is involved in the DD2 peak~\cite{Botha2007May,Adeleye2024Aug}. Therefore, additional theoretical analysis is necessary to gain a better understanding of these materials and their defect structures.

State-of-the-art theoretical approaches mainly consist of density functional theory (DFT) calculations of formation energies and the resulting charge transition levels~\cite{CharacterizationBook2, Freysoldt2014,Alkauskas2016,Kim2020Jul}.
For these DFT calculations, the choice of the right exchange-correlation functional is important. Vidal \textit{et al.}~\cite{Vidal2010} highlighted this through self-consistent GW calculations of the chalcopyrite band structure, and further showed that hybrid functionals provide a satisfactory description of the electronic properties of pristine chalcopyrites. Consequently, most recent studies on the electronic properties of defects in these compounds are based on hybrid functionals, using medium-sized periodic supercells~\cite{ Bailey2010May, Pohl2010Jul, Oikkonen2011Oct, Oikkonen2012Oct, Chen2012, Pohl2013Jun, Bekaert2014Sep,  Han2014Nov, Huang2013Oct, Yee2015Nov,Han2016Nov, Malitckaya2017Jun, Han2017Oct,Shukla2021a,Yang2017Dec, Xiang2023Jun}. 

Most of the defect-related studies on sulfur-based CIGS consider one of the two limiting compounds. For CuInS$_2$ prominent intrinsic point defects have been investigated by Chen \textit{et al.}~\cite{Chen2012}, who also include the O$_\text{S}$ substitutional as a possible defect in non-vacuum growth conditions. Other studies focus on defect complexes; for example, Yang \textit{et al.}~\cite{Yang2017Dec} studied intrinsic defect complexes, while Xiang \textit{et al.}~\cite{Xiang2023Jun} investigated the impact of phosphorus on the doping type. The stability of the different intrinsic defects for different phases of CuGaS$_2$ was studied by Bailey \textit{et al.}~\cite{Bailey2010May}, where the authors demonstrated the native p-type character of the material. Transition metal- and Sn-doped CuGaS$_2$ has been studied by Han \textit{et al.}~\cite{Han2014Nov,Han2016Nov, Han2017Oct} while Pohl \textit{et al.}~\cite{Pohl2010Jul} investigated V$_\text{Cu}$ for four chalcopyrites. Shukla \textit{et al.}~\cite{Shukla2021a} consider both limiting cases and calculate the charge transition levels of a wide range of defects. Based on these, they propose that the Cu$_\text{In}$ and Cu$_\text{Ga}$ antisites are potential defects for the peak D2 (1.1 eV) observed in indium rich CIGS. However, to strengthen their claim, optimization of the HSE parameters (e.g., using the generalized Koopmans' theorem) and a discussion of the limitations of charge transition levels as well as of their connection to optical transitions via lattice relaxation are missing.

The aim of this paper is to provide a complete and accurate study of the defect transition levels in both CuInS$_2$ and CuGaS$_2$ by means of density functional theory within the Heyd, Scuseria and Ernzerhof (HSE) hybrid functional scheme. We limit our study to the isolated intrinsic point defects which are expected to exist based on previous studies of formation energies~\cite{Bailey2010May,Yang2017Dec,Shukla2021a,Xiang2023Jun,Chen2012}. In most previous studies, the mixing parameter~\cite{Yang2017Dec,Xiang2023Jun,Leach2016, Han2016Nov} or the range parameter~\cite{Pohl2010Jul, Chen2012} was adjusted to achieve a band gap close to the experimental value. Han \textit{et al.}~\cite{Han2017Oct} and De\'ak \textit{et al.}~\cite{Deak2017Feb} adjust both parameters. In addition to the band gap, Han \textit{et al.}~\cite{Han2017Oct} used parameters that best describe the Kohn-Sham orbitals of Sn$_\text{Ga}$ within the generalized Koopmans' theorem (gKT).
We follow the idea of De\'ak \textit{et al.}~\cite{Deak2017Feb}, who find the optimal parameter pair for defects in $\beta-\text{Ga}_2\text{O}_3$ by evaluating the parabolicity of partial addition of electrons to neutral V$_\text{O}$. Preferably, this should be done for each defect. However, we investigate a broad defect landscape of both materials, which makes this procedure impractical. We therefore choose a trade-off between computational resources and accuracy by finding the optimal parameter pair for the pristine crystals. We investigate the parabolicity for partial removal of an electron to align with the gKT. To confirm the choice, we calculate the parabolicity of the defects in their neutral charge state.

To further enhance the accuracy delivered in our study, we increase the supercell size to reduce the defect-defect interaction and to provide an updated insight on the charge transition level of the investigated defects.  Compared to the maximum of 144 atom supercells~\cite{Shukla2021a} previously used for charge-transition-level (CTL) calculations, corresponding to a 3$\times$3$\times$1 replication of the conventional 16 atom unit cell, we now employ a 3$\times$3$\times$2 supercell with 288 atoms. This larger supercell increases the minimum defect-defect separation by a factor of 1.48 makes a clear difference for the resulting charge transition levels (CTL).

Moreover, we also deliver a more accurate evaluation of optical transitions associated with defects. This is achieved by considering optical transition levels (OTL) within the total-energy picture.  They can provide us with a more detailed insight into the experimentally observed optical transitions, compared to what CTL alone can deliver. As expected intuitively, for deeper, more localized defects, the lattice contribution to their spectroscopic signature is not negligible.
Accurate calculations of the OTL for the Cu$_\text{In}$ antisite, in its expected charge state, reveal an emission of around 1.1 eV, which aligns well with the peak (D2) observed in photoluminescence measurements.  This confirms previous suggestions that the Cu$_\text{III}$ defect is a suitable candidate, and we also identify V$_\text{Cu}$ and Cu$_\text{i}$ as potential defects for the D1 transition. Furthermore, our results indicate V$_{\text{Cu}}$, Cu$_\text{Ga}$ and Cu$_{\text{i}}$ as possible defects responsible for the DD1 transition in CuGaS$_2$.

The paper is structured as follows. In Section \ref{sec:methods} the methodology for this paper is presented.  In Section \ref{sec:results} we present the charge transition levels of the studied point defects. This is followed by a discussion of the impact of the lattice relaxation on optical transition levels. Finally, we compare these results with photoluminescence measurements, before we conclude our findings in Section \ref{sec:conclusion}.

\section{\label{sec:methods} Methods}

\subsection{Hybrid functional parametrization}
\begin{figure}[b]
\includegraphics{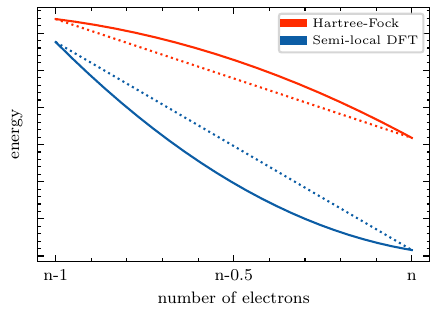}
\caption{\label{fig:semvshf} Scheme of the parabolicity of (semi-)local functionals (blue) and Hartree-Fock (red) for partially removing an electron from the system.  Dotted line indicates the linear behavior if the exact functional was known. (Semi-)local functionals show a positive parabolicity which can favor partially occupied states and delocalize the electron. Hartree-Fock shows a negative parabolicity and favors fully occupied states. Adapted from~\cite{Freysoldt2014}.}
\end{figure}

Hybrid functionals, such as the one introduced by Heyd, Scuseria and Ernzerhof~\cite{Heyd2003},  include to some degree the exact exchange-energy of the Hartree-Fock theory. They partially overcome the well known problem of the band gap underestimation associated with local and semi-local exchange-correlation functionals.  The HSE exchange-correlation energy is given by~\cite{Heyd2003}:

\begin{equation}
 \begin{aligned}
E_\text{XC}^\text{HSE} =&~  \alpha E_\text{X, SR}^\text{HF}(\omega) + (1 - \alpha)  E_\text{X,SR}^\text{PBE}(\omega) \\ 
 &+ E_\text{X,LR}^\text{PBE}(\omega) + E_\text{C}^\text{PBE}~,
\end{aligned}
\end{equation}
which includes a mixing parameter, $\alpha$,  and a range parameter, $\omega$, for the separation of the exchange-energy into a long- and a short-range part. The most commonly used HSE functional is the HSE06 with the parameter pair of $\alpha$ = 0.25 and $\omega$ = 0.2 $\text{\AA}^{-1}$.

This parameter pair is well established for many materials. However, when used for chalcopyrites, the band gap is  underestimated. To study defects in these materials, it is crucial to find a pair of HSE parameters that accurately describes the electronic structure of the system. Consequently, it is necessary to examine two properties of the system to identify the most suitable pair of parameters. Pre-selection is done by comparing the band gaps with the experimental data. These parameters were then further investigated to be compliant with the gKT, in order to obtain an accurate description of the localization of the electronic states.

Compliance with the gKT is achieved with the exact functional, where the total energy must show a piecewise linear behavior with derivative discontinuities at integer occupancies~\cite{Perdew1982Dec}. However, local and semi-local functionals are continuous functions that show a convex parabolic behavior (Figure \ref{fig:semvshf}) with partial removal/addition of electrons. For the Hartree-Fock method, this leads to a concave curvature, because it includes non-local interactions. Within HSE we combine both methods, and a more linear behavior can be achieved by choosing a sensible pair of parameters.

Each defect has different properties, and ideally one should analyze the impact of the parameter pair on the electronic behavior of the respective defects individually. However, this involves relaxations for the investigated defects for different parameter pairs, which is computationally demanding. Instead, we carried out the search for the best parameter pair for the pristine crystals, as a trade-off between computational cost and accuracy. In the investigated range, we find that the best agreement for both compounds is obtained around the mixing parameter $\alpha$ = 0.25. For the range parameter, we find $\omega$ = 0.13 $\text{\AA}^{-1}$ and $\omega$ = 0.10 $\text{\AA}^{-1}$ to satisfy the band gap for the indium and gallium compound, respectively. A detailed description can be found in Appendix \ref{ap:parabol}. The corresponding band gaps for both compounds can be found in Table \ref{tab:bandgaps}. 

\begin{table}[b]
\caption{\label{tab:bandgaps}
Band gap energies (eV) for CuInS$_2$ and CuGaS$_2$. The hybrid functional band gap corresponds to parameters of $\alpha = 0.25$ for both compounds and $\omega = 0.13~\text{\AA}^{-1}$ and $\omega = 0.10~\text{\AA}^{-1}$ for indium and gallium, respectively.}
\begin{ruledtabular}
\begin{tabular}{lcc}
&this work& exp. data\\
\hline
CuInS$_2$ & 1.53 & 1.53\footnote{\label{1sttablefoot}Measurements carried out at 2 K~\cite{Tell1971}}\\
CuGaS$_2$ & 2.52 & 2.53\footref{1sttablefoot}\\
\end{tabular}
\end{ruledtabular}
\end{table}

\subsection{Defect transitions within the total-energy picture\label{different_transitions}}
A common method for studying defects in materials is the supercell approach. When using it, a specific defect is embedded in a finite-size sample of an otherwise perfect structure. In order to be able to draw conclusions about the dilute limit, the supercell needs to be large enough. The stability of a given defect X in its charge state q can then be assessed using the formation energy which is calculated as follows~\cite{ZhangNorthrup,  VanDeWalle1993,Freysoldt2014,Kim2020Jul}: 
\begin{equation}
 \begin{aligned}
  E^\text{f}[X^q] = E_\text{tot}[X^q]  -& E_\text{tot}[\text{bulk}] + \sum_i n_i \mu_i\\
+& q E_\text{F} + E_\text{corr}^q~,  
\end{aligned}  
\end{equation}   
where $E_\text{tot}[X^q]$ is the total energy of the relaxed defective supercell and $E_\text{tot}[\text{bulk}]$ is the total energy of the pristine supercell. The symbol $\mu_i$ denotes the chemical potential of the reservoir of atomic species $i$ of which $n_i$ atoms are added to or removed from the system. The fourth term is the equivalent for the electron reservoir. The chemical potential of the electrons is the Fermi energy, $E_\text{F}$, and $q$ is the number of electrons added to or removed from the defect. This term serves to describe defects in different charge states.

For charged defects, a correction energy, $E_\text{corr}^q$, must be considered in order to account for the spurious interaction between periodically repeated defects. We use the method introduced by Freysoldt \textit{et al.}~\cite{FREYSOLDT,FREYSOLDT2}, where we include the static dielectric tensor ($\epsilon^0 = \epsilon^\text{ion} + \epsilon^\text{el}$). A different approach is necessary to properly capture the correction for optical transition levels, where ionic relaxation is performed only for one of the charge states~\cite{vtl_ref,Falletta2020Jul}. In this case, we only account for the high-frequency value of the dielectric tensor ($\epsilon^\infty \approx \epsilon^\text{el}$).

The formation energy can give us access to important properties of the defect, including the charge transition levels (CTLs)~\cite{Freysoldt2014}:
\begin{equation}
\epsilon(q_1/q_2) = \frac{E^f[X^{q_1}, E_F=0] - E^f[X^{ q_2}, E_F=0]}{q_2 - q_1}~.
\end{equation}
The CTLs correspond to the positions of the Fermi energy at which the two charge states $q_1$ and $q_2$ have the same formation energy.  Under given circumstances (see below), they are comparable with zero-phonon lines (ZPLs) in luminescence. In practice, CTLs are independent of the formation energies and can be calculated as follows:
\begin{equation}
  \epsilon(q_1/q_2) = \frac{[E_\text{tot}[X^{q_1}]+E_\text{corr}^{q_1}] - [E_\text{tot}[X^{q_2}]+E_\text{corr}^{q_2}]}{q_2 - q_1}~.
\end{equation}

In order to investigate the optical properties of defects, ideally one should use many-body approaches. However, such higher-level approaches are computationally expensive and cannot easily be applied to the large supercells that are necessary for defect calculations. Therefore, even for optical properties, one is often constrained to the total-energy picture, defined by the ground state density as the reliable property. In the following, we discuss three approaches for calculating the optical properties on the basis of the total energy. 

The calculation of charge transition levels (CTLs) is the starting point, because they reveal the expected charge state of the defect, at a given Fermi energy, and its type. This can give a first hint at which defects might be responsible for a certain peak. A direct comparison of CTLs with PL peaks can be justified for shallow defects within host materials where the electron-hole interaction is weak.

\begin{figure}[b]
\centering
\includegraphics{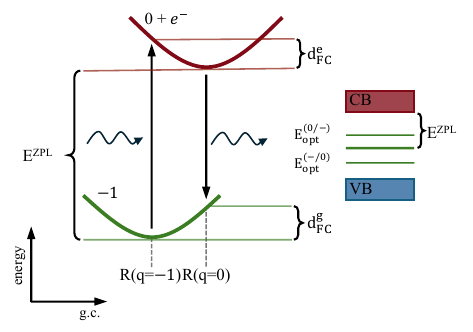}
\caption{\label{fig:OTL_CCD}Illustration of optical transition level (OTL) and charge transition level (CTL) within a configurational coordinate diagram.  In this example, the transition is referenced to the conduction band. The right panel shows a band diagram with the positions of the transition levels. The color code of the left panel corresponds to those of the band diagram (CB = red, defect = green, VB = blue). i) E$^{(-/0)}_\text{opt}$: An electron in the defect state can absorb light, promoting it to the conduction band. This is described by a vertical transition in the atomic ground state (q=-1). ii) E$^{(0/-1)}_\text{opt}$: An electron in the conduction band can transition to a defect state by emitting light. Here the ground state geometry is defined by the neutral charge state. The difference between the vertical transition levels and the charge transition levels corresponds to the Franck-Condon shifts ($\text{d}^\text{e/g}_\text{FC}$), as indicated in the figure.}
\end{figure}

We can improve the comparability by considering that optical (de)excitation is much faster than lattice relaxation. The central idea of this so-called optical transition level (OTL) is that for optical processes the defect geometries do not change instantaneously~\cite{Freysoldt2014, Alkauskas2016}. The calculation is similar to CTL, except that the total energies, $E_\text{tot}[X^{q_1}]$ and $E_\text{tot}[X^{q_2}]$, are both calculated using the geometry of the initial charge sate $X^{q_1}$. We can visualize this within a configuration coordinate diagram (CCD) (Figure \ref{fig:OTL_CCD}). Within the CCD, we can approximate the collective behavior of the atoms in the vicinity of the defect with a harmonic dependence. In this way, it is straightforward to visualize the Franck-Condon shifts for the absorption process (where $E^{(-/0)}_{\text{OTL}} = E_{\text{ZPL}} + d_\text{FC}^e$) and for the emission process (where $E^{(0/-)}_{\text{OTL}} = E_{\text{ZPL}} - d_\text{FC}^g$).

The key advantage of considering vertical transitions is therefore the possibility of capturing the response of the lattice via the Franck-Condon shift. However, within this approach, the electron-hole interaction is not included. A strong interaction would lead to a further reduction of the emission and absorption lines, respectively. This approach is therefore best applicable for defect transitions with small electron-hole interaction, e.g., band-to-defect or defect-to-band transitions involving shallow defects in materials with small exciton-binding energy.

In the third approach, one explicitly conserves the number of electrons in the (de)excitation process. This can be achieved with the $\Delta$-SCF (self-consistent field) approach, sometimes referred to as constrained DFT (cDFT). Here, an excited state can be constructed in terms of single slater determinants, where an electron is constrained into a specific unoccupied Kohn-Sham state. Similarly to the OTL, in this approach we can relax the system in the excited state and use a CCD to analyze and obtain the zero-phonon line and the absorption or emission transitions associated with a given defect. In contrast to the previous two approaches, $\Delta$-SCF  includes electron-hole interactions by construction. However, the convergence of these calculations is delicate as they include a change of electron occupation between discrete defect-related KS states in the band gap and the continuum of states in the valence/conduction band. We therefore focus on the OTLs to discuss the optical properties of the defects investigated in this work.

\section{Results and Discussion\label{sec:results}}

\subsection{Charge transition levels}
\begin{figure*}
\centering
\includegraphics{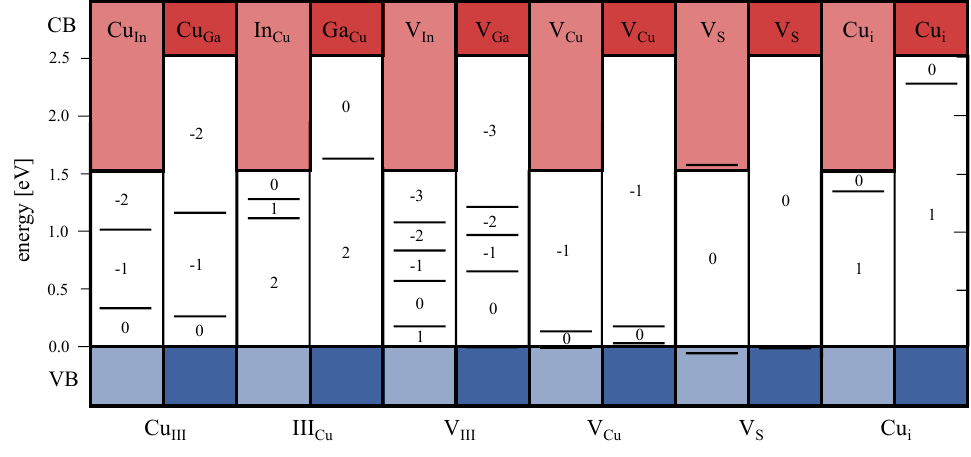}
\caption{\label{fig:CTL}Charge transition levels of the studied intrinsic defects in CuInS$_2$ and CuGaS$_2$.  Defects are grouped by types. Antisites, the group three vacancies and the copper vacancies show deep charge transition levels. The S vacancy has no transition level within the band gap. The Cu$_\text{i}$ has a shows a charge transition level close to the conduction band.}
\end{figure*}
\begin{table}[b]

\caption{\label{tab:CTL}%
Calculated charge transition levels for the most prominent intrinsic defects for CuInS$_2$ and CuGaS$_2$. Energies are in eV.}
\begin{ruledtabular}
\begin{tabular}{l S r l S r}
        CuInS$_2$			        & ~ 			& ~ 	 & CuGaS$_2$			 & ~ 			& ~ 			\\ \hline
        Cu$_{\text{In}}$	        & \phantom{-}0/-1 	    & 0.33   & Cu$_{\text{Ga}}$			             & \phantom{+}0/-1	    & 0.27\\ 
        						& -1/-2 	    & 1.02   & ~ 				     & -1/-2 	    & 1.16\\ 
        ~ 							& ~ 			& ~ 	 & ~ 				     & 	~ 			& ~ 			\\ 
        In$_{\text{Cu}}$	        & +2/+1 		& 1.12   &Ga$_{\text{Cu}}$       & +2/\phantom{+}0		& 1.61\\ 
        ~ 							& +1/\phantom{+}0 		& 1.27   & ~ 					 & ~ 		&  ~\\ 
        ~ 							& ~ 			& ~ 	 & ~ 					 & ~ 			& ~ 			\\ 
        V$_{\text{In}}$		        & +1/\phantom{-}0		    & 0.17   &V$_{\text{Ga}}$	 & +1/\phantom{+}0	    & -0.01			\\ 
        ~ 							& \phantom{-}0/-1    & 0.57   & ~ 					 & \phantom{+}0/-1       & 0.66\\ 
        ~ 							& -1/-2 	    & 0.84   & ~ 					 & -1/-2        & 0.97\\ 
        ~ 							& -2/-3 	    & 1.08   & ~ 					 & -2/-3 	    & 1.22\\ 
        ~ 							& ~			    & ~ 	 & ~ 					 & ~ 			& ~ 			\\ 
        V$_{\text{Cu}}$	            & +1/\phantom{-}0		    & -0.01  &V$_{\text{Cu}}$		 & +1/\phantom{-}0	 		& 0.03\\ 
        ~ 							& \phantom{+}0/-1 	    & 0.14   & ~ 		     & \phantom{+}0/-1 	    & 0.18\\ 
        ~ 							& ~			    & ~ 	 & ~ 					 & ~ 			& ~ 			\\ 
        V$_\text{S}$			    & +1/\phantom{-}0 		    & -0.06  &V$_\text{S}$	 		 & +1/\phantom{-}0 	 	& -0.01\\
        ~ 							& \phantom{+}0/-1 	    & 1.58   & ~ 			 & \phantom{+}0/-1  	    & 2.53\\ 
        ~ 							& ~			    & ~ 	 & ~ 					 & ~ 			& ~ 			\\ 
        Cu$_\text{i}$			    & +1/\phantom{+}0 		& 1.35   &Cu$_\text{i}$          & +1/\phantom{+}0  	& 2.29\\
\end{tabular}
\end{ruledtabular}
\end{table}
\subsubsection{Antisite defects}

Indium and gallium, belonging to the third group (III), have two more valence electrons than copper; this means that when the latter sits in a group-III site it should behave as an acceptor. Both Cu$_\text{In}$ and Cu$_\text{Ga}$ have two charge transition levels from 0 to -2 at similar positions with respect to the valence band maximum, confirming this behavior (Figure \ref{fig:CTL} and Table \ref{tab:CTL}). 

In the case of gallium and indium sitting on the copper site, our results show deep transition levels. Increasing the Fermi energy changes the charge state from +2 to 0, which is characteristic for a donor. The CTLs for In$_\text{Cu}$ are clearly separated by a few hundred meV, whereas for Ga$_\text{Cu}$ we observe no stable +1 charge state and a direct transition from +2 to 0.

The comparison of our results with previous studies is not straightforward, as they were carried out with different supercell sizes and different HSE parameters. However, the results identify similar CTLs at different positions within the band gap.  Unlike the other studies, Xiang \textit{et al.}~\cite{Xiang2023Jun} find the Cu$_\text{In}$ to be a shallow acceptor without any neutral charge state within the band gap. Furthermore, for In$_\text{Cu}$, Chen \textit{et al.}~\cite{Chen2012} find a relatively shallow CTL $\epsilon$(+2/+1), while Yang \textit{et al.}~\cite{Yang2017Dec} and Xiang \textit{et al.}~\cite{Xiang2023Jun} do not observe any charge transition level with a stable charge state of In$^{+2}_\text{Cu}$. In contrast, the other studies reveal a deeper position of the $\epsilon$(+2/+1) CTL.

\subsubsection{Vacancy defects}
Both group III (In and Ga) vacancies show three charge transition levels ranging from 0 to -3 at similar positions with respect to the valence band. While this results in acceptor behavior for the V$_\text{Ga}$,  V$_\text{In}$ is an amphoteric defect with an additional CTL $\epsilon$(+1/0) relatively close to the valence band. The V$_\text{Cu}$ exhibits a single CTL $\epsilon$(0/-1) located around 0.14 eV and 0.18 eV, respectively. In contrast to all the other defects investigated in this study, the sulfur vacancy does not exhibit any charge transition level within the band gap, and the neutral charge state is the most stable.

The most significant difference from previous studies is the amphoteric behavior of V$_\text{In}$, a phenomenon absent in the other studies~\cite{Shukla2021a,Chen2012,Yang2017Dec}. However, the analysis of the charge difference for +1 and 0 shows a very localized behavior of the charge, indicating a stable charge state. The results of Pohl \textit{et al.}~\cite{Pohl2010Jul} indicate a shallow V$_\text{Cu}$ without any charge transition levels in both compounds, while the other works that discuss V$_\text{Cu}$ have a $\epsilon$(0/-1) transition level relatively close to the valence band. Another contrast is the $\epsilon$(+2/0) for the V$_\text{S}$ deep within the band gap in the study by Chen \textit{et al.}~\cite{Chen2012}, a defect that has not been studied by the other articles. 

\subsubsection{Copper interstitials}
To calculate the properties of the copper interstitial, we assume a behavior analogous to that previously observed in selenide-based chalcopyrites, where the atomic ground state is located in the center of the cation octahedron~\cite{Pohl2013}. We find that a Cu$_\text{i}$ defect exhibits a charge transition level $\epsilon$(+1/0) at similar positions relative to the conduction band in both compounds. In contrast, Yang et al~\cite{Yang2017Dec} and Xiang \textit{et al.}~\cite{Xiang2023Jun} report that Cu$_\text{i}$ acts as a shallow donor without exhibiting any distinct charge transition levels.

\subsection{Optical transition levels}
\begin{table}
    \centering

    \begin{ruledtabular}
    \begin{tabular}{lcc|lcc}
    \textbf{CuInS$_2$}    &transition& OTL  & \textbf{CuGaS$_2$} &transition& OTL  \\\hline
    CBM as ref. & ~&~&~& ~&~ \\
    Cu$_{\text{In}}^\text{em}$&$(0\rightarrow -1)$ & 1.12 &
    Cu$_{\text{Ga}}^\text{em}$&$(0\rightarrow -1)$ & 2.20 \\
    V$_\text{In}^\text{em}$&$(+1\rightarrow 0)$  & 1.24 &
    V$_\text{Ga}^\text{em}$&$(0\rightarrow -1)$  & 1.75 \\
    V$_\text{Cu}^\text{em} $&$(0\rightarrow -1)$   & 1.36 &
    V$_\text{Cu}^\text{em}$&$(0\rightarrow -1)$   & 2.27 \\
    \hline
    VBM as ref. & ~&~&~& ~&~ \\
    In$_{\text{Cu}}^\text{abs}$&$(+2 \rightarrow +1)$   & 1.39 &
    Ga$_{\text{Cu}}^\text{abs}$&$(+2 \rightarrow +1)$   & 2.26 \\
    In$_{\text{Cu}}^\text{em}$&$(+1 \rightarrow +2)$   & 0.74 &
    Ga$_{\text{Cu}}^\text{em}$&$(+1 \rightarrow +2)$   & 0.45 \\
    Cu$_\text{i}^\text{abs}$&$(+1 \rightarrow 0)$    & 1.36  &
    Cu$_\text{i}^\text{abs}$&$(+1 \rightarrow 0)$   & 2.29 \\
    Cu$_\text{i}^\text{em}$&$(0 \rightarrow +1)$    & 1.31 &
    Cu$_\text{i}^\text{em}$&$(0 \rightarrow +1)$   & 2.24 \\
    \end{tabular}
    \end{ruledtabular}
    \caption{Optical transition levels (in eV). The charge states in the “transition” column denote the initial and final charge states of the defect. The three defects listed in the upper part have an empty Kohn-Sham state in the band gap, which allows an optical transition from the conduction band into the defect level, thereby charging the defect negatively. In contrast, III$_\text{Cu}$ and Cu$_\text{i}$ do not exhibit an empty Kohn-Sham state in the band gap and their transitions are referenced to the valence band, as they must absorb an electron before emission. }
    \label{tab:otl}
\end{table}

\begin{table}
    \centering
    \begin{ruledtabular}
    \begin{tabular}{lcc|lcc}
    \textbf{CuInS$_2$}    &transition& d$_\text{FC}$  & \textbf{CuGaS$_2$} &transition& d$_\text{FC}$ \\\hline
    CBM as ref. & ~&~&~& ~&~ \\
    Cu$_{\text{In}}^\text{em}$&$(0\rightarrow -1)$ & 0.08 &
    Cu$_{\text{Ga}}^\text{em}$&$(0\rightarrow -1)$ & 0.06 \\
    V$_\text{In}^\text{em}$&$(+1\rightarrow 0)$  & 0.12 &
    V$_\text{Ga}^\text{em}$&$(0\rightarrow -1)$  &  0.12\\
    V$_\text{Cu}^\text{em} $&$(0\rightarrow -1)$   & 0.03 &
    V$_\text{Cu}^\text{em}$&$(0\rightarrow -1)$   & 0.08 \\
    \hline
    VBM as ref. & ~&~&~& ~&~ \\
    In$_{\text{Cu}}^\text{abs}$&$(+2 \rightarrow +1)$   & 0.27 &
    Ga$_{\text{Cu}}^\text{abs}$&$(+2 \rightarrow +1)$   & 0.62 \\
    In$_{\text{Cu}}^\text{em}$&$(+1 \rightarrow +2)$   & 0.38 &
    Ga$_{\text{Cu}}^\text{em}$&$(+1 \rightarrow +2)$   & 1.17 \\
    Cu$_\text{i}^\text{abs}$&$(+1 \rightarrow 0)$    & 0.01 &
    Cu$_\text{i}^\text{abs}$&$(+1 \rightarrow 0)$   & 0.00 \\
    Cu$_\text{i}^\text{em}$&$(0 \rightarrow +1)$    & 0.04 &
    Cu$_\text{i}^\text{em}$&$(0 \rightarrow +1)$   & 0.05 \\
    \end{tabular}
    \end{ruledtabular}
    \caption{Franck-Condon shift (in eV) for the calculated optical transition levels. The charge states in the “transition” column denote the initial and final charge states of the defect. The three defects listed in the upper part have an empty Kohn-Sham state in the band gap, which allows an optical transition from the conduction band into the defect level, thereby charging the defect negatively. In contrast, III$_\text{Cu}$ and Cu$_\text{i}$ do not exhibit an empty Kohn-Sham state in the band gap and their transitions are referenced to the valence band, as they must absorb an electron before emission.}
    \label{tab:dfc}
\end{table}

\begin{figure}
\includegraphics{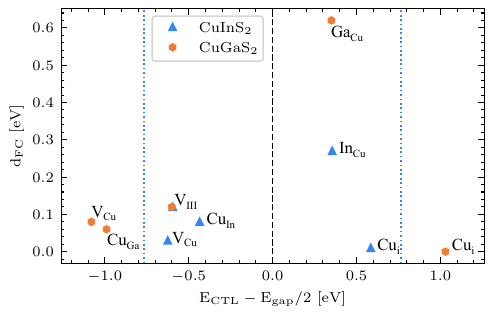}
\caption{\label{fig:dfc} Franck-Condon shift as a function of the charge-transition level (CTL) position relative to midgap. CTLs closer to midgap induce larger lattice distortions upon changing the charge state. Blue dotted vertical lines mark the band edges of CuInS$_2$, and the x-axis limits correspond to those of CuGaS$_2$.}
\end{figure}

CuInS$_2$ and CuGaS$_2$ are native p-type semiconductors with the Fermi level located close to the valence band~\cite{Bailey2010May, Xiang2023Jun}. This determines the most stable charge states of the defects and the starting point for subsequent investigations, including optical transition levels.

Both possible emission processes involve a different initial condition for the defect. For a transition from the conduction band to the defect, empty Kohn-Sham states within the band gap are necessary, whereas for a transition from the defect to the valence band a defect state within the band gap has to be occupied. The first condition is fulfilled for the studied acceptors and the V$_\text{In}$, for which we can consider a conduction band to defect transition as depicted in Figure \ref{fig:OTL_CCD}. The other defects (III$_\text{Cu}$ and Cu$_\text{i}$) in their preferred charge state only show a defect related state once ionized. Therefore, we have to consider a transition from the defect to the valence band, following the ionization of the defect (absorption).

The emission energies obtained from the optical transition levels and the corresponding Franck-Condon shifts are listed in Table \ref{tab:otl} and Table \ref{tab:dfc}, respectively. The first number in parentheses relates to the initial charge state, and the second number to the final charge state.

To estimate the predictive power of CTL when used to study the optical properties of defects, the correlation between the position of CTL and the corresponding d$_\text{FC}$ is visualized in Figure \ref{fig:dfc}. As expected, the shifts are small for defects with a CTL close to the band edges (Figure \ref{fig:dfc}), however, not negligible.

Our results highlight that if we want to compare charge transition levels to optical measurements, the Franck-Condon shifts cannot be neglected. While pristine chalcopyrites show weak electron-hole interactions~\cite{Gil2012}, this may not hold for transitions involving deep defects, where additional corrections are expected. Nevertheless, we compare our results with existing measurements, as they improve on earlier studies that considered only CTLs. For transition levels closer to the respective bands, it is reasonable to compare the OTLs with optical measurements.

\subsection{Connection to experiments}
\begin{figure*}
\includegraphics{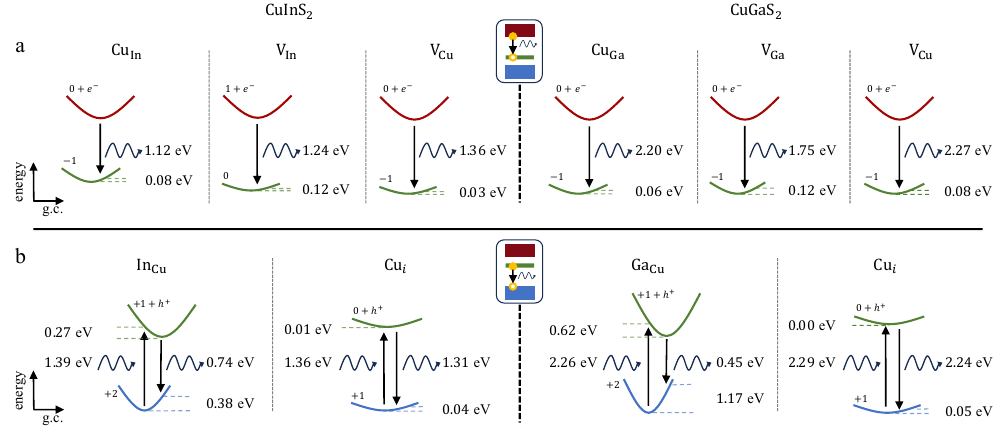}
\caption{\label{fig:otl}Sketch of the optical/vertical transition levels within a configuration coordinate diagram. The color code of the parabolas describes the position of the electron and is consistent with the inset pictures (CB = red, defect = green, VB = blue). \textbf{a} corresponds to transitions from the conduction band to the defect, while in \textbf{b} the transition is between defect and valence band. Left displays to defects in Defects in CuInS$_2$ and right in CuGaS$_2$.}
\end{figure*}

The optical transition levels give additional information on the transitions observed for defects in both sulfide-based chalcopyrites compared to the charge transition levels calculated in previous studies. Although this is still an approximate treatment of the transitions, it is natural to compare these with photoluminescence measurements.

Photoluminescence measurements performed on the chalcopyrite Cu(Ga,In)S$_2$~\cite{Shukla2021a} reveal two defect-induced transitions at around 1.35 (D1) and 1.1 eV (D2). These measurements were performed on samples with a [Ga]/[Ga+In] ratio ranging from 0.12 to 0.18 and a band gap of approximately 1.58 eV. In the same study, this transition has been attributed to the Cu$_\text{III}$ antisites, as they exhibit a CTL at this energy and the photoluminescence peak becomes more prominent under copper-rich growth conditions, reflecting the observation for measurements with different copper content. 

The present study investigates intrinsic defects in the two limiting chemical compositions of Cu(In,Ga)S$_2$. The calculated charge transition levels for the indium-based chalcopyrite reveal five potential candidates for the observed spectroscopic signatures. In particular, for the D1 transition (1.35 eV), both the V$_\text{In}$ and V$_\text{Cu}$ vacancies, as well as Cu$_\text{i}$, exhibit charge transition levels close to the experimentally observed peak. However, the Franck-Condon shift for V$_\text{In}$ shifts the optical transition to lower energies around 1.24 eV, making it less likely to be responsible for the peak. Formation energy calculations support both other defects as candidates for this peak \cite{Yang2017Dec, Shukla2021a}.

For the D2 transition, we find Cu$_\text{In}$ as a possible candidate with an optical transition level of 1.12 eV. This closely matches the experimentally observed peak at 1.1 eV and aligns with the previous study by Shukla \textit{et al.} \cite{Shukla2021a}. The second candidate for the D2 transition is the In$_\text{Cu}$ defect which has a charge transition level at 1.12 eV above the valence band maximum, suggesting that it could be involved in the 1.1 eV transition. However, analysis of optical transition levels indicates a substantial Franck-Condon shift of more than 0.27 eV for both absorption and emission lines, which would rule out this defect (Figure \ref{fig:otl}).

Studies on CuGaS$_2$ also show two deep defect transitions around 2.12-2.18  eV (DD1) and 1.80-1.85 eV (DD2)~\cite{Eberhardt2003Sep,Botha2007May, Adeleye2024Aug}. The DD1 is broad peak which is highly composition dependent, and it is suggested that more than one defect might be involved in the transition~\cite{Botha2007May,Adeleye2024Aug}. We find that three defects exhibit an optical transition level that aligns with the measurements for the DD1 peak in CuGaS$_2$: V$_{\text{Cu}}$, Cu$_{\text{Ga}}$ and Cu$_{\text{i}}$ (Figure \ref{fig:otl}).  Assuming a similar behavior of the defect formation energies for Cu$_{\text{i}}$ as observed in the indium compound, all three defects have low formation energies for either of the growth conditions in the experiments~\cite{Bailey2010May,Yang2017Dec}. This could support the hypothesis that more than one defect is involved in the observed peak.

Finally, our results show only one possible candidate in the range of the DD2, namely V$_\text{Ga}$. However, this defect has a high formation energy in the stoichiometric phase \cite{Bailey2010May}, which aligns with the high formation energy of V$_\text{In}$ \cite{Yang2017Dec}, making it an unlikely candidate.

Although these calculations offer an updated view on the optical transitions, they still involve many approximations and have to be used with caution in comparison with photoluminescence experiments.

\section{\label{sec:conclusion}Conclusion}
In summary, we have carried out a thorough search to find a well suited HSE parameter pair for chalcopyrites, by identifying a pair that shows the experimental band gap and also complies with the generalized Koopmans' theorem. The search has been done for the pristine material as a trade-off between computational cost and accuracy. We found good agreement for the indium compound for $\alpha$ = 0.25 and $\omega$ = 0.13 $\text{\AA}^{-1}$ and for the gallium compound for $\alpha$ = 0.25 and $\omega$ = 0.10 $\text{\AA}^{-1}$.
Additionally, we have increased the supercell size  to 288 atoms compared to previous studies using smaller cells with a maximum of 144 atoms~\cite{Shukla2021a}. 

We have further calculated the corresponding charge transition levels for common intrinsic point defects, namely Cu$_\text{III}$,  III$_\text{Cu}$, V$_\text{III}$, V$_\text{Cu}$, V$_\text{S}$, and Cu$_i$, where III stands for group three elements:  indium and gallium. We discuss the differences with previous studies and the main distinction is the amphoteric behavior of V$_\text{In}$ which is observed only in our study.

The relevance of both compounds as a p-type absorber material for solar cells motivated the investigation of the optical properties. We discuss the difference between different total-energy methods used to obtain optical properties. Previous reports on these materials use the charge transition levels directly to explain the origin of the deep defect signals. 

In the present work, we assume that optical transitions are vertical transitions within the configuration coordinate diagram, thereby accounting for the different timescales associated with electron excitation and ionic relaxation. To the best of our knowledge, this approach has not yet been applied to chalcopyrites. We therefore shed new light on the optical properties of the defects, making them better comparable to photoluminescence measurements. In agreement with theory, our results reveal that for deeper charge transition levels, the contribution of ionic relaxation is significant, as evidenced by a Franck-Condon shift of several hundred meV. We also observe that for charge transition levels close to the respective bands edges the shift is relatively small, making them suitable for initial comparisons with photoluminescence measurements.

Specifically, we observe that a defect-related  transition (D1)  in CuInS$_2$ might be related to V$_\text{Cu}$ or Cu$_\text{i}$, as they exhibit an optical transition level comparable to the experimental data. Furthermore, we identify Cu$_\text{In}$ as a candidate for a second transition in close proximity to the peak, labeled D2, in photoluminescence measurements performed on indium-rich Cu(In,Ga)S$_2$. As a limiting case for the quaternary compound, this suggests Cu$_\text{III}$ as a good candidate, supporting previous results. For CuGaS$_2$, we identified candidates for a deep transition around 2.15 eV. The broad defect peak and its strong compositional dependence suggest that multiple defects are involved~\cite{Botha2007May,Adeleye2024Aug}, which aligns well with the calculated optical transition levels of V$_{\text{Cu}}$, Cu$_\text{Ga}$ and Cu$_{\text{i}}$.

\section{Data availability}
VASP input files, calculation outputs, and Python scripts are available at https://doi.org/10.5281/zenodo.17910868.

\section{Acknowledgments}
This research was funded by the Luxembourg National Research Fund (FNR), grant reference PRIDE17/12246511/PACE.
Fruitful discussions with Susanne Siebentritt are gratefully acknowledged. We also acknowledge the use of the HPC facilities of the University of Luxembourg~\cite{VCPKVO_HPCCT22}.

\appendix
\section{Calculation details}

The \textit{ab-initio} calculation were performed with the VASP~\cite{vasp1, vasp2, vasp3,vasp_paw} software package. The cutoff energy for the plane waves converged at 400 eV. The forces for the convergence of $1 \cdot 10^{-3}$ eV/\text{\AA} and an electronic self consistency criterion of $1\cdot 10^{-6}$ eV were used. For all calculations we used the PBE exchange correlation functional and the projector augmented-wave method (PAW).

For the parabolicity calculations, the conventional cell was pre-relaxed with PBE before relaxing the structure for different HSE parameters. Since this includes many calculations, these calculations were performed with a 2$\times$2$\times$1 Monkhorst-Pack grid to pre-screen the band gaps. We increased the mesh density to 4$\times$4$\times$2 to find a range of band gaps on which to perform parabolic calculations.

For atomic relaxation, the defect and its nearest neighbors were displaced from their equilibrium positions. Initially, a pre-relaxation step was carried out using PBE with fixed supercell dimensions and without enforcing symmetry constraints, employing an atomic convergence criterion of $1\cdot 10^{-3}$ eV/\text{\AA} and an electronic convergence criterion of $1\cdot 10^{-5}$ eV. Subsequent relaxations using the HSE functional included symmetry constraints and used a relaxed convergence criterion of $1\cdot 10^{-1}$ eV/\text{\AA} for atomic positions and $1\cdot 10^{-4}$ eV for electronic convergence. 

The dielectric constant was obtained with a Monkhorst-Pack grid of 3$\times$3$\times$2 for the conventional unit cell and the default electric force field of 0.01 eV/\AA . For the indium compound, the diagonal elements of the purely electronic dielectric tensor are $[\epsilon^{el}_{xx},\epsilon^{el}_{yy},\epsilon^{el}_{zz}] = [6.12,6.12,6.17]$. The ion contribution to the dielectric tensor shows slightly anisotropic behavior with $[\epsilon^{ion}_{xx},\epsilon^{ion}_{yy},\epsilon^{ion}_{zz}]  = [2.66,2.66,3.46]$. Likewise, for the gallium compound, $[\epsilon^{el}_{xx},\epsilon^{el}_{yy},\epsilon^{el}_{zz}] = [5.97,5.97,5.98]$ and $[\epsilon^{ion}_{xx},\epsilon^{ion}_{yy},\epsilon^{ion}_{zz}]  = [2.34,2.34,3.14]$.

\section{Hybrid functional parameterization}\label{ap:parabol}
To find the best parameter pair for the hybrid functional, band gap calculations were performed around the standard HSE06 parameter pair ($\alpha$: 0.2-0.3,  $\omega$: 0.1-0.3 $\text{\AA}^{-1}$). Here, a sparse Monkhorst-Pack grid was used to find the trends and to calculate the number of parameter pair candidates substantially. 
We evaluated the parabolicity of CuInS$_2$ for selected parameter pairs. The exact exchange-correlation functional exhibits a piecewise linear behavior, featuring derivative discontinuities at integer electron numbers. Therefore, to improve the performance of the hybrid functional, we identify parameter pairs that closely approximate linearity. 

\begin{figure}[!t]
\includegraphics{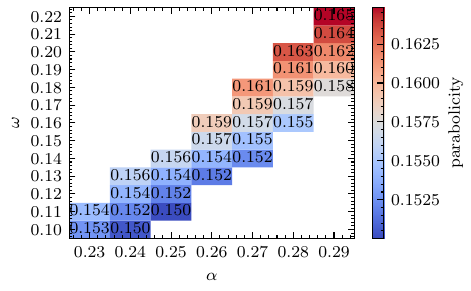}
\caption{\label{fig:parabolicity_in} Parabolicity results for CuInS$_2$ for parameter pairs that are in the range of previous band gap calculations.}  
\end{figure}

\begin{figure}[!t]
\includegraphics{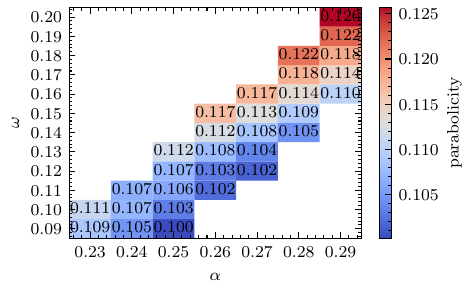}
\caption{\label{fig:parabolicity_ga} Parabolicity results for CuGaS$_2$ for parameter pairs that are in the range of previous band gap calculations.} 
\end{figure}

For the parabolicity calculations we keep the geometry fixed. We identify the trends for CuInS$_2$  and CuGaS$_2$ for a Monkorst-Pack grid of 4$\times$4$\times$2. We calculate the total energies in five steps from N to N-1 electrons (Figure \ref{fig:parabolicity_in} and Figure \ref{fig:parabolicity_ga}). Although the overall parabolicity is relatively high compared to previous studies focusing on a single defect \cite{Deak2017Feb}, we find a local minimum around the standard mixing parameter and linearity is best described around the standard HSE mixing parameter $\alpha=0.25$. Subsequent band gap calculations were performed using the same Monkhorst-Pack density used for supercell calculations, and we find that the best agreement with the experimental band gaps is obtained with parameters $\alpha=0.25$ and $\omega$ = 0.13 $\text{\AA}^{-1}$ and $\omega$ = 0.10 $\text{\AA}^{-1}$, respectively.

We have investigated the transferability to defects, by partial addition and removal of acceptors and donors in their neutral charge state, respectively. This makes sure that we investigate a defect related Kohn-Sham orbital. The calculations include the charge correction for the vertical transition and are displayed in Table \ref{tab:defect_parabol}. The parameter pair result in small parabolicities for the defects in their neutral charge state, indicating that they are sufficiently well described with the parameters obtained from pristine calculations.

\begin{table}
    \centering
    \begin{ruledtabular}
    \begin{tabular}{lr|lr}
    \textbf{CuInS$_2$}    &parabolicity & \textbf{CuGaS$_2$}&parabolicity \\\hline
    Cu$_\text{In}^0$ & 0.021 &
    Cu$_\text{Ga}^0$ & 0.071\\
    In$_\text{Cu}^0$ & 0.077&
    Ga$_\text{Cu}^0$ & 0.104\\
    V$_\text{In}^0$ & 0.076&
    V$_\text{Ga}^0$ & 0.021\\
    V$_\text{Cu}^0$ & 0.091&
    V$_\text{Cu}^0$ & 0.065\\
    Cu$_\text{i}^0$ & 0.075 &
    Cu$_\text{i}^0$ & 0.050\\
    \end{tabular}
    \end{ruledtabular}
    \caption{Parabolicity of the total energy for the defects in their neutral charge state, calculated for partial addition and removal for acceptors and donors, respectively. The calculations include the charge correction for vertical transition.}
    \label{tab:defect_parabol}
\end{table}

\newpage

\end{document}